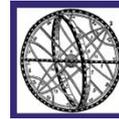

# Image of Inverted World Tree on the Stone Slab and Vessels of the Bronze Age

Larisa N. Vodolazhskaya[1], Pavel A. Larenok[2], Mikhail Yu. Nevsky[3]

[1] Southern Federal University (SFU), Rostov-on-Don, Russian Federation;
E-mails: larisavodol@aaatec.org, larisavodol@gmail.com
[2] NP "Yuzharheologiya", Rostov-on-Don, Russian Federation; E-mail: dao2@inbox.ru
[3] Southern Federal University (SFU), Rostov-on-Don, Russian Federation; E-mails:
munevsky@sfedu.ru

## Abstract

The article presents the results of the study of petroglyphs on a unique stone slab discovered near the kurgan 1 of the kurgan field Varvarinsky I (Rostov Oblast, Russia). Analysis of features of the location and style of petroglyph "tree" showed that the branches could determine semi-minor semiaxes m of the "dial" ellipses of analemmatic sundials with semi-major axis M = 24.2 cm for medium and high (northern) latitudes up to the North Pole and "tree" marks the direction to the North. The desire to construct a tool with which it was possible to build a "dial" for analemmatic sundial for different latitude up to the latitude of the North Pole could appear under the influence of the dominant mythological ideas about sacred geography of the surrounding world in which the North is the World Mountain, World Tree or the abode of the gods. In the article the authors conclude that the petroglyph "tree" on Varvarinsky slab marks the direction to the North as a projection of the North Pole of the world on the earth's surface, which simulates the slab. Petroglyph is an image of an inverted World Tree, when viewed from the gnomon / man. The trunk of the Tree corresponds to the astronomical world axis, and its branches symbolize the visible daily path of the Sun and, perhaps, the night paths of stars across the sky at different latitudes. During the research was carried out a comparative analysis of the petroglyph "tree" with images of the inverted tree on the Srubna vessels. According to the analysis it was concluded that a mythical World Tree of Srubna population is likely to be a tree of pine family. Sign of the inverted tree on the Srubna vessels has been interpreted by as a symbol of the direction of the North and / or the north pole of the world; and accompanying signs were interpreted as symbols of circumpolar stars and constellations (asterisms).

## Introduction

In 2013 archaeologist A.V. Fayfert discovered the stone slab with petroglyphs in Sholokhov district of Rostov Oblast (Russia) (Fig. 1). The slab was found in 30 m to the southeast of kurgan 1 of the kurgan field Varvarinsky I (49.5º N, 41.4º E). Kurgan has a height of 0.4 m at the moment, and it is pulled on the North-South line. Plate weight around 70 kg. The second side of slab does not contain petrogliphs [1, p. 27-28].

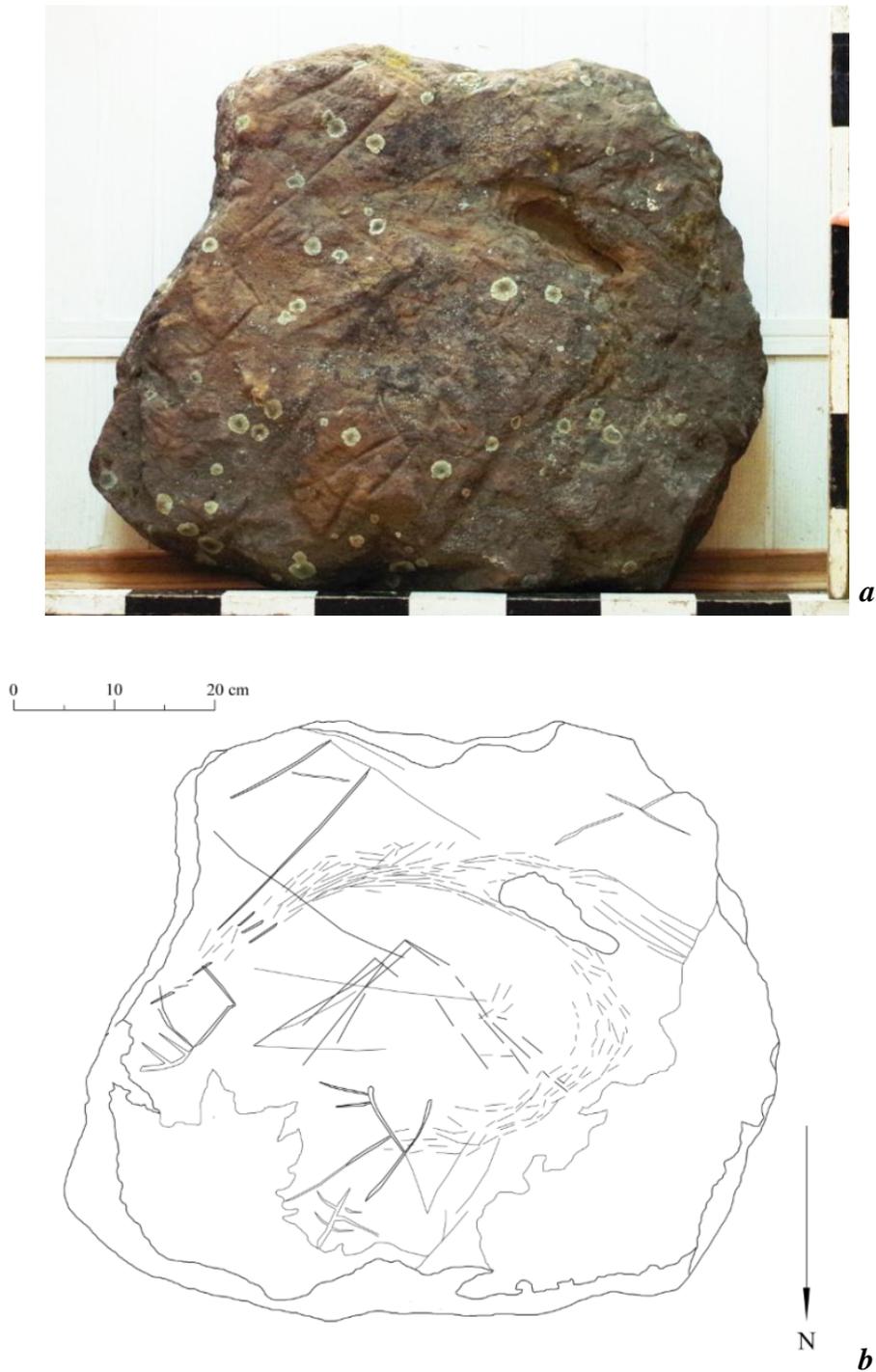

**Figure 1.** Kurgan field Varvarinsky I, kurgan 1 (neighborhood), a stone slab with petroglyphs: *a* - photo of the slab[1] (by A.D. Vodolazhskiy, 2014), *b* – drawing of the slab (by L.N. Vodolazhskaya, 2015). Slab is worth perpendicular to the floor (the alignment was carried out using plumb lines, slab supported by special fasteners located between the slab and the wall) [2].

---

[1] The photo was taken with the camera, Pentax K-50. For reduction of distortions the process of photographing of vertically standing (checked by means of a plumb) slab produced at a distance about four meters. The lens of the camera, directed perpendicular to slab surface, was opposite to its center. The resulting slab image occupied a small part in the center of the frame. Location photometers monitored by building levels. The flash is not used.

An analysis of the petroglyphs on the slab from the vicinity of the kurgan field Varvarinsky I was carried out with the help complex methods of natural science, which already shown to be effective in the interdisciplinary research of ancient structures and artifacts [3-14].

This analysis led to the conclusion that it is the prototype of Srubna slab from burial of Kurgan field Tavriya-1 [15, p. 62; 16, p. 4-14; 17]. Petroglyph in the form of an elliptical groove in the center of the Varvarinsky slab by its parameters correspond to elliptical the location of wells - hour marks of analemmatic sundial on the Tavria-1 slab. However, there are no wells in the Varvarinsky slab, which could serve as hour marks, therefore it is impossible consider it as a sundial. It is only a prototype of analemmatic sundial - slabs with petroglyphs from Srubna burials from Rostov and Donetsk regions, and precedes them [2]. Varvarinsky slab dates from the Bronze Age in a fairly wide range, starting from the end of IV millennium BC. To date its more precisely almost impossible, since slab discovered without archaeological context. However, since Varvarinsky slab is the prototype of slabs with petroglyphs from Srubnaya burials, then as one of the boundaries of its dating it is possible to consider the late Bronze Age or XVII-XII centuries BC.

In previous studies able to establish that Varvarinsky slab was made, in the first place, as an example for the markup of the "dial" ellipse of analemmatic sundial with semi-major axis M≈20.5 cm, for latitude ≈40º N and ≈46º N [2].

On the Varvarinsky slab "north of" the main ellipse, beaten out in the form of a groove in the center of slab, easily visible petroglyph "tree" (Fig. 2).

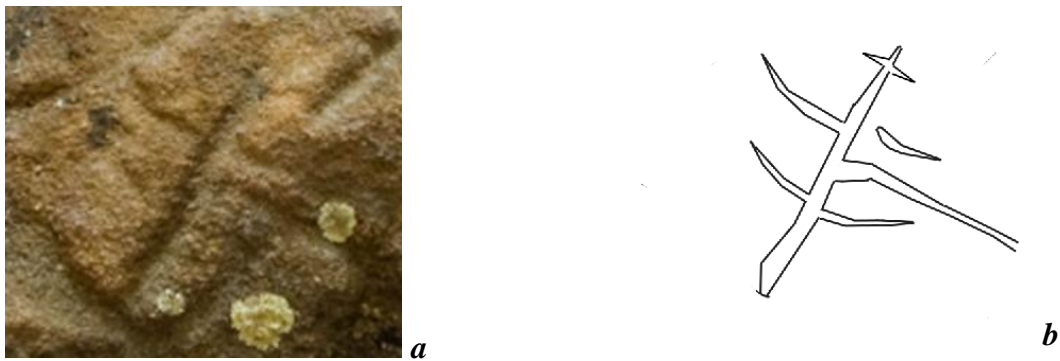

Figure 2. Kurgan field Varvarinsky I, kurgan 1 (neighborhood), a stone slab with petroglyphs: *a* - part of the photo with a "tree", *b* - drawing of petroglyph "tree".

"Tree" is projected on the Y-axis in the range of about $y \in [18.2; 24.3]$ (cm), which corresponds to a range of small semiaxis from m≈18.2 cm to m≈24.3 cm. In building on drawing the ellipse with a semimajor axis M = 20.5 cm, it was found that the top of the tree corresponds to an ellipse with the small semiaxis m≈19 cm (Fig. 3).

Ellipse with a semimajor axis M = 20.5 cm and small semiaxis m≈19 cm will correspond to latitude φ≈68ºN - close to the latitude of the Arctic Circle, equal 66.57º N. In this case, even when the minor axis will have a maximum length equal to m=M≈20.5 cm only a small part of the petroglyph "tree" will be used - its apex. All petroglyph "tree" or the majority of it can be involved only when the semi-major axis M is greater than 20.5 cm.

As has been discovered in the previous study, the length of the semi-major axis M on the slab defines petroglyph "rhombus". Although that basically Varvarinsky slab was intended for marking ellipses with the length of the semimajor axis M≈20.5 cm, defined by center of "rhombus", existed the possibility of marking ellipses with a semimajor axis defined by the closest to the center of the slab angle of "rhombus" with M≈17.5 cm. Similarly, it would be possible to try to analyze the possibility of marking ellipses and with the help of the far corner of "rhombus", which specifies the length of semi-major axis of the ellipse M≈24.2 cm, which exceeds 20.5 cm.

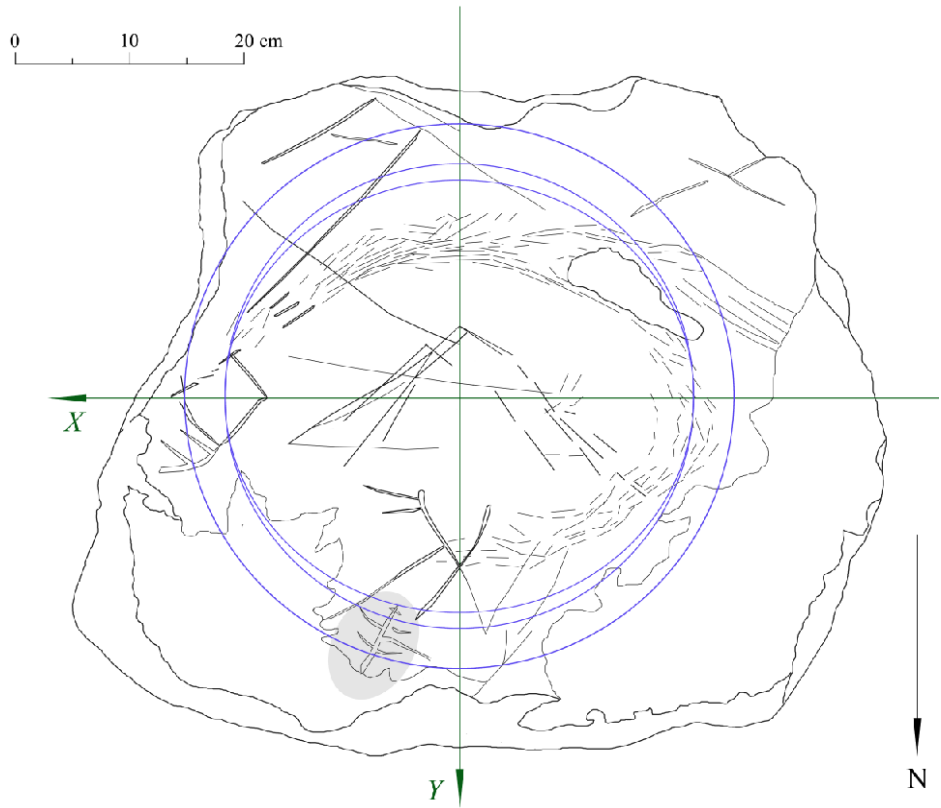

**Figure 3.** Kurgan field Varvarinsky I, kurgan 1 (neighborhood), a stone slab with petroglyphs. The inner circles of blue color - ellipses of analemmatic sundial "dial" with M≈20.5 cm and m≈19 cm for latitude ≈68º N and m=M≈20.5 cm for latitude 90º N (North Pole). The outer circle - ellipse of "dial" with m=M≈24.2 cm to the latitude of the North Pole. Petroglyph "tree" is highlighted gray oval.

Performed preliminary calculations showed that for ellipses with M≈24.2 cm distance Z, for which it is necessary to move the gnomon in the days of the solstice, will roughly correspond to the distance at which branchings is located, associated with petroglyph of "rhombus". The measured distance of these branchings: 4.2 cm, 5.0 cm, 6.1 cm, 6.5 cm [2].

Latitude φ for analemmatic "dial" with a semimajor axis M≈24.2 cm with known Z can be calculated by the formula 1:

$$\varphi = \arccos\left(Z \Big/ \left(M \cdot tg\,\delta_{ss}\right)\right) \tag{1}$$

where φ – latitude of locality, Z – distance at which shifts the gnomon at the summer solstice, M – semi-major axis of the ellipse, $\delta_{ss}=\varepsilon$ – the declination of the Sun at the summer solstice.

The results of calculations by the formula 1 are presented in Table 1.

**Table 1.** The calculated values of the latitude φ with M = 24.2 cm for the given values of gnomon shift Z.

| | φ, ° |
|---|---|
| $M, cm$ / $Z, cm$ | 24.2 |
| 6.5 | 52.61 |
| 6.1 | 55.26 |
| 5.0 | 62.15 |
| 4.2 | 66.90 |

The lengths of the small semi-axes have been calculated by the formula 2:

$$m = M \cdot \sin \varphi \qquad (2)$$

where m − small semi-axes of the ellipse, M − semi-major axis of the ellipse, φ − latitude of locality.

Calculation results are given in Table 2.

**Table 2.** Calculated values of small semi-axes of the ellipse m with M = 24.2 cm for the given values of the latitude φ.

| | m, cm |
|---|---|
| $M, cm$ / $\varphi, °$ | 24.2 |
| 52.61 | 19.3 |
| 55.26 | 20.0 |
| 62.15 | 21.5 |
| 66.90 | 22.3 |

Ellipses, built for M = 24.2 cm and for the calculated small semi-axes m, were marked on the drawing of Varvarinsky slab (Fig. 4).

In Figure 4 shows that all ellipses associated with branching points of petroglyph "rhombus" and pass through the branchings of petroglyph "tree" (Fig. 4). Not involved only the most distant from the center of coordinates branching − lower branches of the "tree." Around them can be associated with an ellipse − circle of "dial", constructed for the latitude of the North Pole, for which m=M≈24.2 cm.

Analysis of the values of calculated latitudes φ revealed their approximate equality latitudes φ$_{ph}$, on which are observed the phenomenon of qualitative changes in natural light during the night near the summer solstice (Table. 3).

As the table shows 3, mean difference (average error) between the calculated latitudes and latitudes closest to them on which the phenomenon of qualitative changes in natural light at night near the summer solstice are observed, Δφ$_{av}$≈182 km. However for the branching point with Z=19.2 cm margin of error is ≈444 km and significantly exceeds the error for all other latitudes. If we exclude the markup for Z=19.2 cm, the average error will be ≈95 km, which corresponds to about 1° of latitude. According to our estimates most marks were made with the same precision on Varvarinsky slab [2].

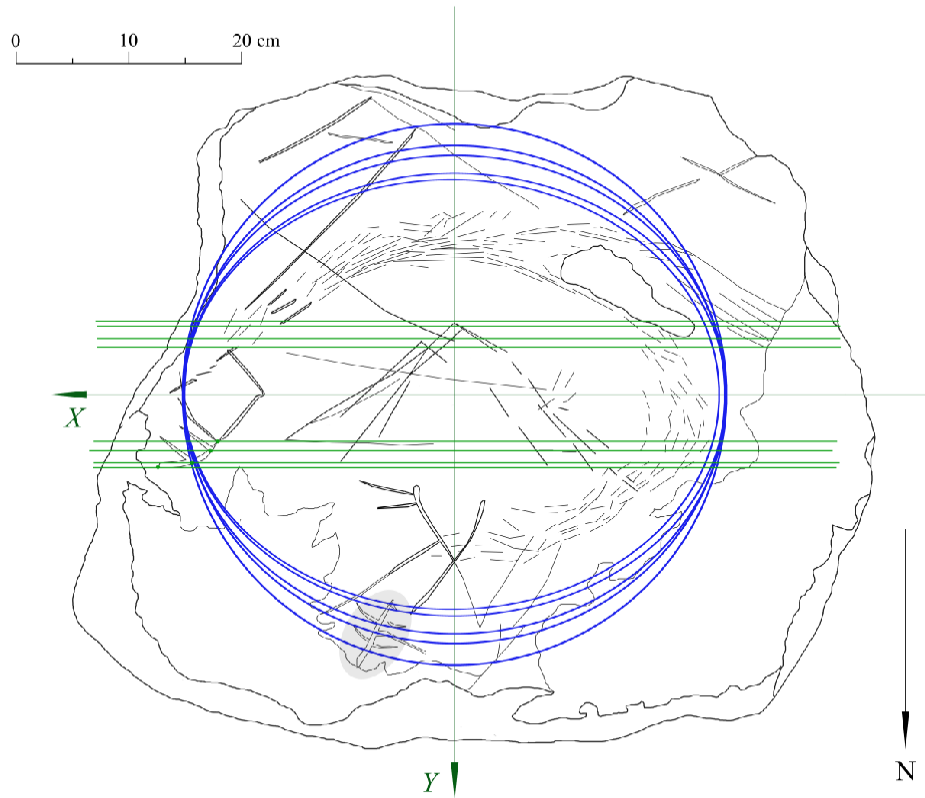

**Figure 3.** Kurgan field Varvarinsky I, kurgan 1 (neighborhood), a stone slab with petroglyphs, drawing of slab with deposited ellipses of the analemmatic sundial "dials" with semi-major axis M≈24.2 cm. Green lines, passing through the points of petroglyph "rhombus" branchings, note the distance Z, on which gnomon is necessary to shift in the days of solstices. Petroglyph "tree" is highlighted gray oval.

Thus, petroglyph "tree" may define minor semiaxes of the "dial" ellipse of analemmatic sundials m with M=24.2 cm for northern latitudes up to the North Pole, and marks on the slab the direction to the north like the arrow on modern maps and drawings.

Discovered on Varvarinsky slab auxiliary markings for construction of ellipses of analemmatic sundials on the northern latitudes is not evidence the reality of traveling to the North in the Bronze Age. It is the most probable that it shows idea of the ancient population of how design features of the analemmatic sundials for different territories had to change. Data for northern territories, up to the North Pole, could be received by means of qualitative (not connected with calculations) extrapolations – distribution of the conclusions received from observation of visible motion of the Sun at the southern and middle latitudes to northern latitudes. Than to more northern latitudes ellipse is marked, the minor semiaxes *m* is less differs from semi-major axis *M*. At the North Pole they must be equal and the ellipse will be transformed into a circle.

The aspiration to construct a tool for the latitude of the North Pole and reach the pole could occur in the ancient population, for example, under the influence of the dominant mythological ideas about sacred geography of the surrounding world, about the location of the abode of the gods.

**Table 3.** Calculated values of north latitude φ for an "dial" ellipse of analemmatic sundial with semi-major axis M = 24.2 cm for different values of small semi-axes m.

m − small semi-axes, *φ* − latitude of locality, φ$_{ph}$ − latitude at which is observed phenomenon of a qualitative change lighting at night near the summer solstice, Δ*φ* − difference between the calculated latitude and closest to it latitude of the observed phenomenon of qualitative alteration of the natural lighting at night near the summer solstice *(Δφ=φ-φ$_{ph}$), Δl* − distance corresponding to the difference between the calculated latitude and the nearest latitude of the observed phenomenon.

| *m*, cm | *φ*, ° | *φ$_{ph}$*, ° | Δ*φ*, ° | Δ*l*, km | Phenomenon |
|---|---|---|---|---|---|
| 24.2 | 90.0 | 90.0 | 0.0 | 0.0 | The latitude of the North Pole, where in the summer polar days with duration of more than six months are observed |
| 22.3 | 66.90 | 66.57 | 0.33 | 35.9 | The latitude of the Arctic Circle, beginning with which and to the north in the summer polar days (when the Sun does not descend below the horizon longer than one day in the year) are observed |
| 21.4 | 62.15 | 60.57 | 1.58 | 174.0 | The latitude of "white nights", beginning with which and to the north, near the summer solstice the Sun does not descend lower than 6º (occur only civil twilight) and the starry sky is impossible to observe |
| 19.8 | 55.26 | 54.57 | 0.69 | 75.6 | The latitude, beginning with which and to the north near the summer solstice the Sun does not descend lower than 12º (occur only civil and nautical twilight) and only the brightest stars are observed |
| 19.2 | 52.61 | 48.57 | 4.04 | 444.2 | The latitude, beginning with which and to the north near the summer solstice the Sun does not descend lower than 18º (come all kinds of twilight, but the astronomical night does not occur), so that the observation of the Milky Way and faint stars is impossible |

So modern researchers define tribes of Srubna culture as Indo-Iranian ethnic groups [18] or Iranian group of Indo-European language family [19-22]. An idea about the astronomical and natural phenomena specific to the latitude of the Far North existed among the ancient Iranians. For example, in the description of abode of the first king Yima the following phrases are present: "The stars, Moon and Sun seem setting and rising one time", "year seemed one day" (Videvdat 2.40-41). In the ancient Indian sacred books and epics reported about the constellations which do not descend below the horizon and called "nonsetting"; polar star[2] "in the middle of heaven"; distant countries, where day and night lasts for half a year, etc. [23, p. 76]. Such descriptions are characteristic for circumpolar regions.

---

[2] Various stars are becoming polar stars - the stars closest to the Pole of the World - because of the precession of the equinoxes and proper motions of stars at different times.

In the ancient Indian and Iranian traditions of the world mountain, which is located in the North, often appears. So the in the oldest parts of the "Avesta", created at the end of the II millennium BC, the direction to the North associated with the sacred mountain Hara Berezaiti, around which hosts of heaven revolved. Mountain was situated, as well as the Indian mythical Mount Meru, at far North and also, like Meru, is the abode of the gods. The Avesta describes that "blissful abode" of Yima is near Hara Berezaiti [24, p. 46-48], where the chosen people live among the beautiful nature, without diseases, worries and envy, where heat, cold, death there were not (Yasht 10.50-51).

Legends about of northern country Hyperborea was first mentioned by Homer in the Iliad (Iliad 15. 171; 19. 358) and described in detail by Herodotus (Herodot History, Book IV, Chapters 32-36) are also likely to reflect the mythical representation about a happy northern country, similar to the ancient Indian and Avestan traditions. In ancient sources described what people in Hyperborea lived to old age accompanied by songs, dances, music and feasts, in the eternal fun and prayers (*Pind. Pyth.* X 29-4T; *Plin. Nat. hist.* IV 26).

"World Pillar", "Axis for Peace" (Axis mundi), "The World Tree" (Arbor mundi) and the like, which are usually placed in the sacred center of the world are cultural and historical versions of the World mountains [25, p. 398-406].

Petroglyph "tree" on Varvarinsky slab marks the direction to the North, as a projection of the North Pole of the World on the surface of the Earth, which simulates by the slab. Petroglyph is the image of the World Tree / World Mountains in fact. The trunk of the "tree" can be regarded as accordance to the astronomical Axis of the World[3]. And its branches can be seen as symbols of the Sun visible daily paths across the sky at different latitudes. Petroglyph "tree" on the slab somewhat displaced relative to the Y-axis - the direction to the North in the plane of the slab (Fig. 6a). Perhaps it was an attempt display the fact that the Pole of the World[4] and, accordingly, the polar star, at all latitudes, except the North Pole, does not coincide with zenith ("center" of the sky). Polus of the World will always be slightly below it (Fig. 6b), except for the "ideal" the direction to the North, observed only at the North Pole, where the North Pole of the World coincides with the point of zenith. A similar way depicted, for example, the wheels in the planning projections, when they were drawn beside a cart in the Bronze Age sometimes (Fig. 6c) [26].

Gnomon of analemmatic sundial, which is a vertically standing rod, is a symbol and model of a standing person (Fig. 6b). If look at the petroglyph "tree" on the part of the gnomon / person, the "tree" would be inverted - with roots to the edge of slab and the top - to the center and the gnomon (Figure 6a). Petroglyph "tree" is the projection of Pole of the World / polar star on the surface of the slab in fact and symbolizes the World Tree, located in the sky with the apex directed towards the ground and the roots directed to the sky.

The image of an inverted World Tree existed in various traditions, starting with the ancient Indian sacred texts, the Arab citations to Plato, writings of Philo of Alexandria, and ending Manichean and Kabbalistic sources, ancient Russian literature and Great Russianin incantations [28]. In India, most often as the World Tree is seen inverted tree - Asvattha (Ficus religiosa) (Atharva Veda II, 7, 3): " The root is at the top, branches are at the bottom, it is - the eternal ashvatta" (Katha Upanishad II, 3, 1) or "Its root is at the top - three feet Brahman (its) branches - space, air, fire, water, earth and more. This is Brahman, calling a single Asvattha" (Maytrayaniya Upanishad VI, 4). In the Russian incantations the following description of an inverted World Tree is found: "On the sea on Okeyane, on an island on Kurgan white birch tree stands, branches are down, roots are up" [29]. "Shamanic tree", which has an inverted view, was found on the river Botchi in the settlements of Orochi also [29]. Birch Lapps used to create an image of the god of thunder, and, the body is made from its trunk, and the head was made from the roots [30, p. 165]. The Lapps did the image of the god of thunder from birch, torso made of its trunk and the head is made from the roots [30, p. 165]. Description of the image of an inverted tree in indigenous shamanic traditions of the northern Siberian and Far Eastern peoples can also be found and in studies of ethnographers V.V. Tcharnolussky [31, p. 79], L.Y. Sternberg [32], Ch.M. Taksami [33].

---

[3] Axis of the World - an imaginary line passing through the Center of the World, around which there is a rotation of the celestial sphere.

[4] Pole of the World - the point on the celestial sphere, around which there is a visible daily motion of stars due to the Earth's rotation around its axis.

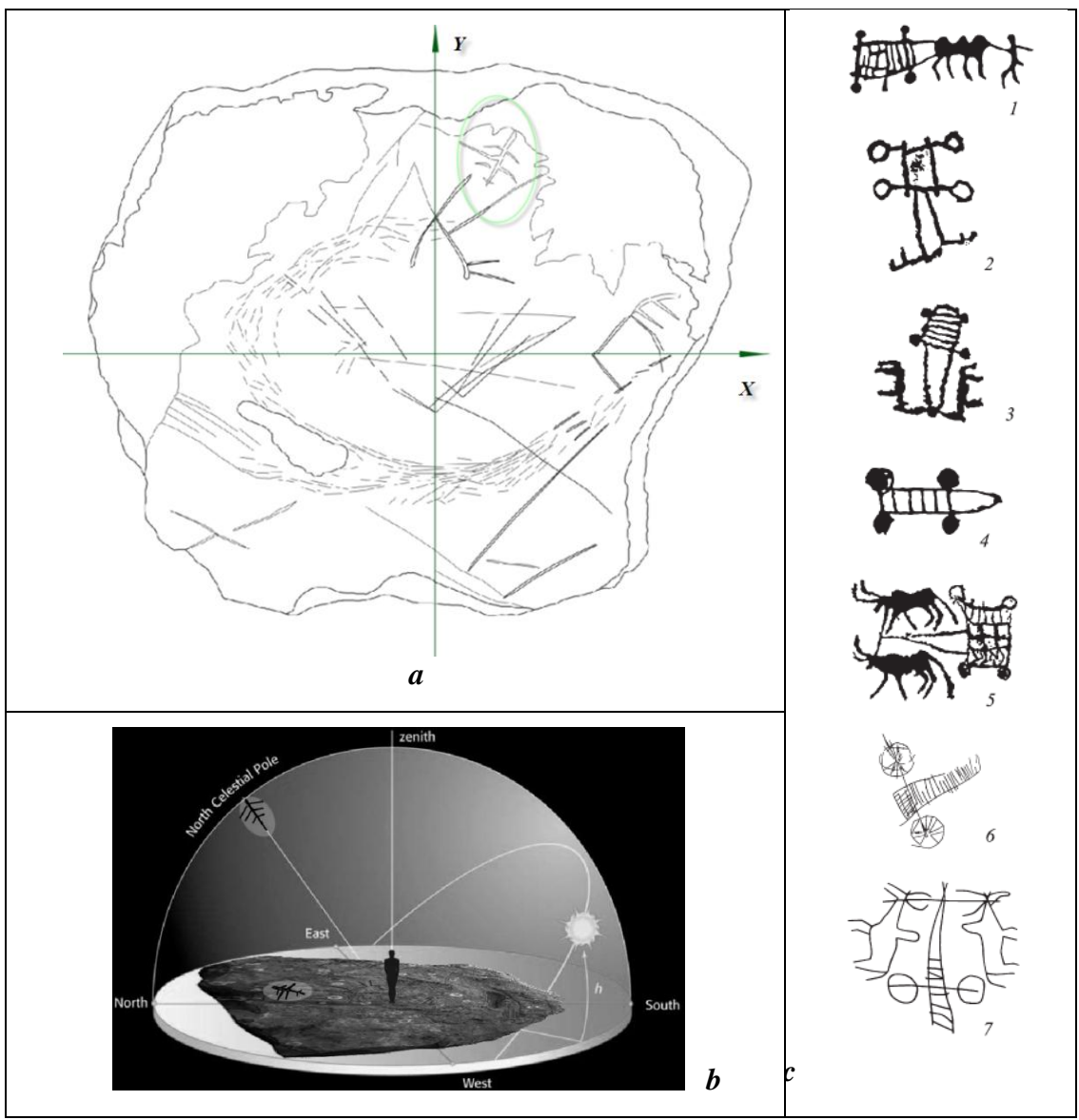

**Figure 6.** Kurgan field Varvarinsky I, kurgan 1 (neighborhood), a stone slab with petroglyphs: ***a*** – drawing, drawing slab; petroglyph "tree" enclosed in an oval green color, ***b*** – modeling of the earth's surface using Varvarinsky slab (figure is based on [27, fig 1.91.]), ***c*** – image of wheeled carts of the Bronze Age in the planned projection [26, Fig. 19].

The reason of the inverted location of the World Tree / shamanic tree in these traditions is not explained. Taking into account that the image of the inverted World Tree meets in traditional cosmologies much less frequently than the image is not of the inverted tree, but it is present in the most ancient written sources - the Vedas, and was discovered by us on a stone slab of the Bronze Age, we think that this fact is associated with a larger antiquity of the inverted image of the world tree, compared to non-inverted. Appears, that throughout the ages, has occurred gradual loss of knowledge about the rational component of this symbol, and image of the inverted tree has been replaced by a more ordinary and understandable image - the image is not inverted tree. About initial image is preserved only fragmentary evidence, scattered among different cultures.

Thus a rational explanation of the inverted image of the World Tree, in our view, due to its symbolic identification with the Axis of the World. In this case the base of its stem (root) can be considered as the North Pole of the World, or the polar star, and the branchesas are visible paths of the Sun and perhaps the stars at different latitudes (or at different distances from the North Pole of the world or the geographic North Pole).

On the Srubna of vessels – molded pottery - inverted tree image encountered, too (Fig. 7a – 7d). In some cases, next to the inverted tree inverted animal is portrayed as on the vessel from the burial ground Fedorovsky II (Fig. 7a). In Figure 7e vessel with an image unturned trees or shrubs shows, and in Figure 7f - with image unturned animals, most likely horses, shows for comparison.

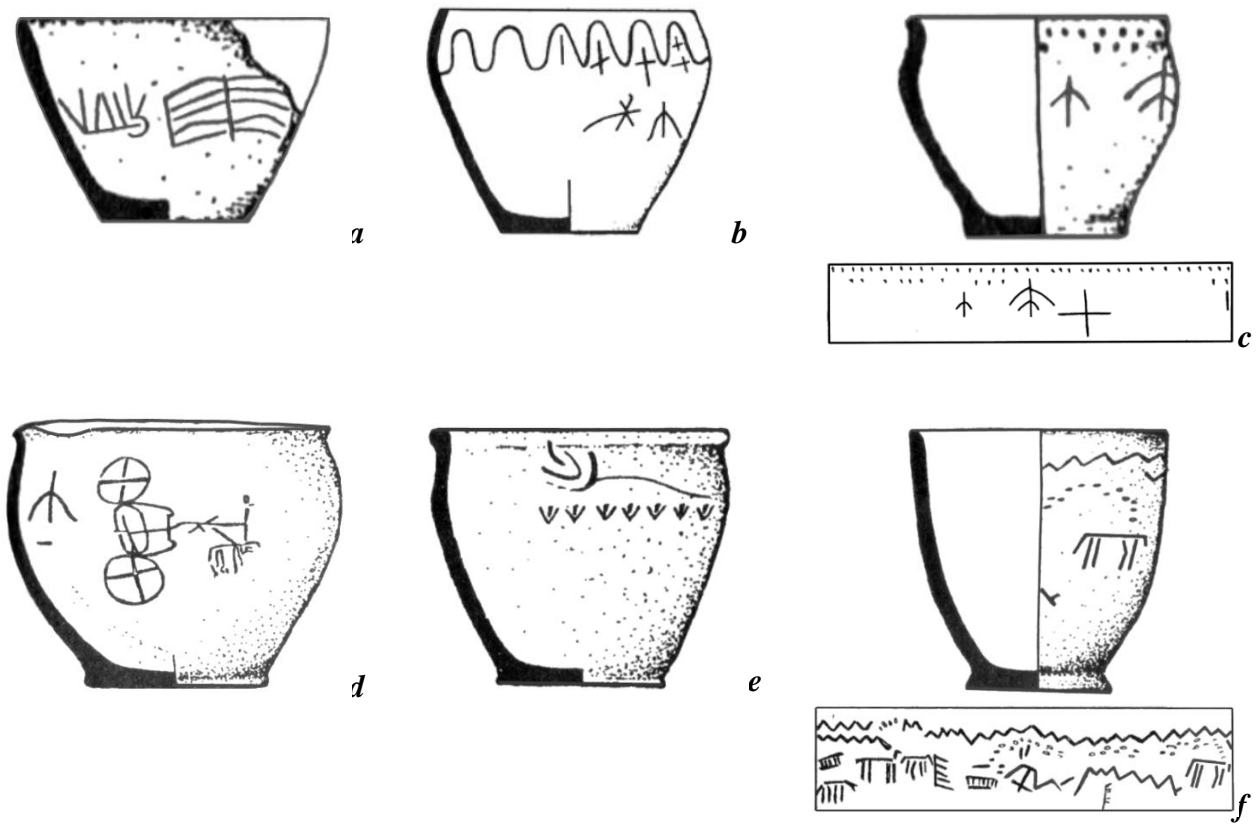

**Figure 7.** Srubna vessels with signs: *a* - burial ground II Fedorovsky, kurgan 3, burial 3 [34, Fig. 58.3], *b* - burial ground Staraya Toyda, kurgan 4, mound [34, Fig. 4.1], *c* - burial ground Bol'shaya Dmitrievka II, kurgan 5, burial 3 [34, Fig. 4.7], *d* - burial ground Sukhaya Saratovka, mound 2, burial 2 [34, Fig. 26.7], *e* - burial ground Zamozhnoe, kurgan 2, 3 burial [34, Fig. 32.3], *f* - burial ground Polyanki [34, Fig. 42.4].

In cases of realistic image of the tree (Fig. 7a, 7c), discrepancies in the interpretation of the signs of tree does not arise. On both images tree branches are arranged symmetrically with respect to the trunk and edges of branches raised toward the top of the tree. In the first case this feature is less pronounced and the branches are arranged almost horizontally, as in the case of tree on Varvarinsky slab. The pyramidal shape of the crown and whorled branching are characteristic of coniferous trees of pine family. And, pines, have cone-shaped krone in youth, and in old age - roundish or umbrella-shaped, flat krone (Fig. 8). Trees of family of pine (Pinaceae) are eurysynusic in many regions: from tropical to subarctic geographical belts, but mainly, in moderate belt of the northern hemisphere [35, page 133-173]. Thus, as mythical World tree the srubny population, most likely, represented tree of family of pine. This or that sort of this family could serve as local symbol: Pine, Fir-tree, Cedar, Larch, Fir, the most widespread in the specific region.

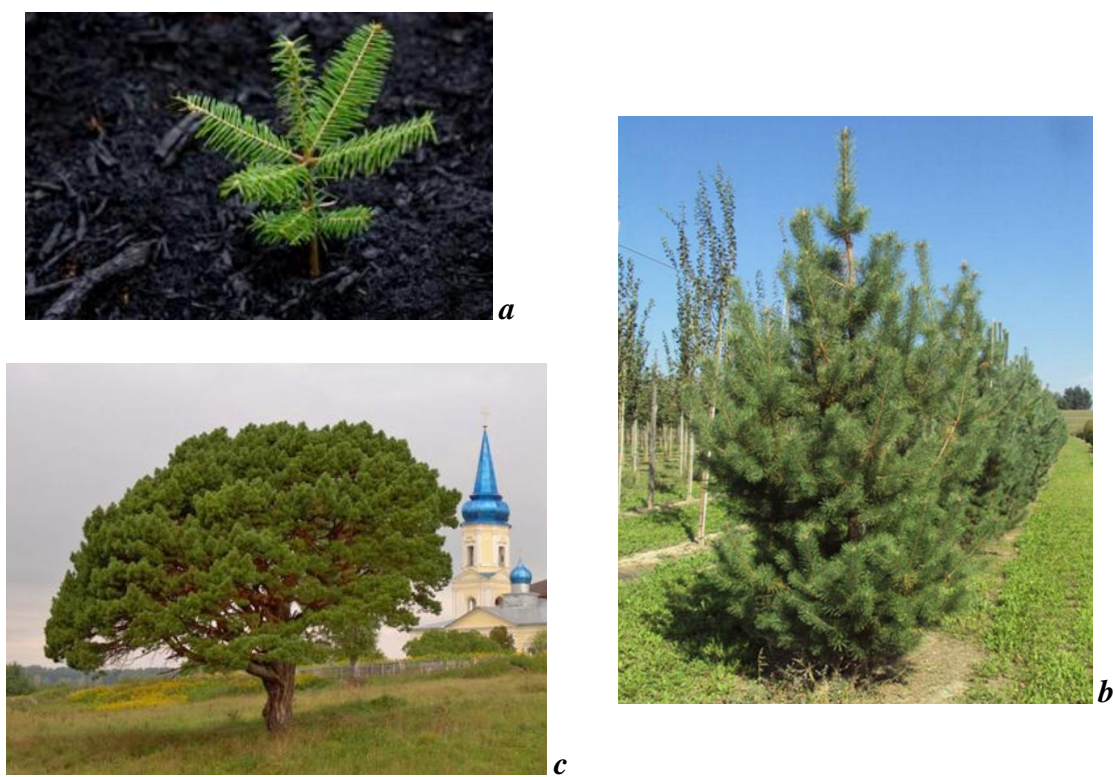

**Figure 8.** Family Pine (Pinaceae), genus Pine (Pinus sylvestris): *a* - saplings[5]; *b* - young trees[6]; *c* - tree of age over 300 years[7].

In cases more abstract images of a tree - signs in the form of two branches on one stalk (Fig. 7b, 7c, 7d), a sign that we interpret as the schematic representation of tree, some researchers are sometimes described as anthropomorphic sign [34, p. 64; 36, p. 48-50], as a sign of boom or a flying bird (the signs in the form of an acute angle with a long bisector) [37].

However, from our point of view, the interpretation of such signs as symbols of tree, looks more justified, especially when you consider that in the described cases, we talking about the inverted World Tree associated with the world of gods and deceased ancestors. In ethnography and folklore of many peoples there are ideas that in the realm of the dead all the way around - upside down. Ritual inversion practiced and carriers Sruba culture. Most brightly it is demonstrated posed upside down vessels from altars and burials [38].

---

[5] http://hvoinie.ru/wp-content/uploads/2012/08/Посадка-сосны-обыкновенной-осенью.jpeg
[6] http://www.deepdale-trees.co.uk/trees/2011/03-Pinus-sylvestris.html
[7] http://www.geocaching.su/?pn=101&cid=6858

At Varvarinsky slab petroglyph "tree" is, in our view, a symbol of the World Tree located in the area of North Pole of the World and in the plane of the slab indicates the direction to the geographic North Pole, as a projection of the North Pole of the World on the earth's surface, that simulates the slab. Therefore, Srubna images of inverted tree we also propose to consider, as a symbols of the North Pole of the World and the geographic North Pole. In Figure 7b six-pointed star is drawn next with a schematic view of an inverted tree, in our view, represents the polar star, although some scholars interpret it as the image of the bull [36]. Four-beam stars are depicted closer to the mouth of the vessel. It seems that they represent a group of stars (the constellation or asterism – readily distinguishable group of stars) in the vicinity of the pole star. In Figure 7c inverted tree is depicted next to the four-beam star in a similar manner. Above - a strip of the points located along the corolla, from our point of view, depicts the starry sky. Symbolic image of the inverted tree, but smaller sizes, is near inverted tree. We assume that in such cases (doubles image of inverted tree) a second, smaller size tree, located farther away from the "polar star", can be a symbol of the zenith point. Than starry sky observation latitude will be closer to the north, the polar star (the North Pole of the World) and the zenith point will be located closer to each other and they practically will coincide at latitude of the North Pole.

North Star was Beta Ursae Minoris (β Ursa Minor) during the existence of Srubna culture, therefore inverted animal in the picture next to inverted tree in Figure 7a, can be regarded as the constellation Ursa Minor and / or Ursa Major, which are located close to each other. In the ancient Iranian tradition the North was venerated in connection with the location of the constellation Ursa Major, which was perceived as the main constellation in the sky [39, p. 80]. Therefore, standing next to inverted tree animal in Figure 7a may also symbolize the constellation Ursa Major. It is known that this constellation called Ursa Major (Hyginus 2, 1.1) or Woz (Hyginus 2, 2.2) in Ancient Greece already. However, for the first time under the "bearish" name Αρκτος this constellation is found only in Homer (Odyssey V, 262-281). In the Rig Veda plural rkshās "stars" (Rig Veda I, 24, 10) is used for the name of this constellation, and the constellation called the "Seven Stars" [40, p. 581]. It is therefore unlikely that the bear is depicted in Fedorovsky vessel.

In Mesopotamia the two constellations associated with carts. Ursa Major was called the "Great Cart" and Ursa Minor - "Heavenly Cart" or "Cart of Anu", where Anu - the supreme god of heaven and father of the gods in the Mesopotamian pantheon [41, p. 295, 301]. The carts were harnessed by oxen in Mesopotamia, as a rule, but the horns are clearly distinguishable in images usually. Animal of Fyodorovsky vessel does not have horns. On a vessel from the burial ground Sukhaya Saratovka (Fig. 7d) chariot drawn by, is likely, to a horse is shown next to the symbol of an inverted tree. We assume, that this image of "Heaven carts" of the supreme god of the sky of Srubna pantheon.

In the ancient Indian mythology Dyaus was the god of the sky, he was personified sky and father of the gods, including Adityas - heavenly gods, originally (AB I, 32, 4). In the Rig Veda, he is always mentioned along with Prithvi – goddess of the earth and mother of the gods (RV I, 160). Fatherhood of Dyaus is his only personified a trait almost, and its zoomorphic images - a bull (RW I 160, 3; V 36, 5; 58, 6) or a stallion (X 68, 11) [42, p. 202]. However, since the beginning of the Vedic period, place of Dyaus was occupied by another god - Varuna, who kept all the heavenly attributes. Varuna is the greatest of the gods of the Vedic pantheon (RV II 27, 10; V 85, 3; VII 87, 6; X 132, 4). He is considered paramount among the heavenly gods - Aditi. Varuna is associated with cosmic waters in all their diversity, he is the guardian of truth and justice, "Pantocrator," the king above the world, gods and humans (RV II 27, 10; V 85, 3; VII 87, 6; X 132, 4) . Varuna the one who created the world and keeps it (PB IV 42, 3; VIII 41, 5). He lights up the sky and the earth, strengthens the Sun, measures the Earth with the help of the Sun, picks it up at the sky, his clothes are day and night. Varuna gave the movement to the Sun; it is his eye (I 50, 6), he himself is a thousand-eyes (VII 34, 10).

However, the heavenly chariot in the Indo-European traditions is often considered as a solar chariot. Mitra refers to solar deities in ancient Indian tradition, but in early Vedic period Varuna forms a pair with Mithra, which is regarded as something single - Varuna-Mitra. Varuna along with Mitra models a space as a whole. But Varuna constantly associate with the night, sometimes was related with the day. Varuna "paved the way for the Sun" and made a "great channels for days" (PB VII 87, 1). Mitra also had more pronounced solar functions and was not associated with the night. Mitra was in the inside of the frame indicated by the kingdom of Varuna; he - like the golden sun embryo in cosmic waters [43, p. 596].

Varuna in the Veda also described as god of the sky, which supports an inverted World Tree:

> "In the bottomless (space) King Varuna with the pure force of action
> Keep straight top of the tree.
> (Branches) are directed downwards. Their foundation - on top.
> Let will root in us the rays!
> Because the king Varuna, made a wide
> Way for the Sun, order to follow (behind him)..." [44, p. 28]
>
> (RV I, 24, 7-8)

Thus, the chariot on the vessel from Sukhaya Saratovka, close to the inverted tree, most likely, is the image of the heavenly chariot of the supreme god is analog of Varuna. The horse is depicted in Fedorovskoye vessel near the tree, perhaps symbolizing the heavenly chariot. In the the Vedic language aśvattha ( "World Tree" - ficus religiosa), aśvayupa ( "sacrificial pillar") formed by aśva-, means "horse" [45, p. 75-138].

In the Russian folklore Ursa Major constellation is sometimes called the "horse to the hitching post" or "humpbacked gelding", that some researchers explain the Turkic influence [46, p. 84-89].

It should also be noted that this constellation was depicted as a bull's front legs originally in Egypt, and later as a whole the figure of the bull; Babylonians called the Ursa Major constellation – "Cart", the ancient Romans – "Seven bulls" and the Greeks – "Woz". "Elk" or "deer" designation of the Ursa Major constellation are common in northern Eurasia [47], [48, p. 100-124]. The two stars of Ursa Minor, one of which is the Polar Star, called White and Grey Gelding among Kazakhs [49, p. 12-14]. In a number of Russian dialects the Polar Star is called "Stozhar", around which elk walks or horse on hitching post [50, p. 326]. Chukchi considered that Polar Star - fixed stake for tether reindeers , around which other stars are moving [51, p. 112].

Interpretation of the inverted images as symbols of the heavenly deities, occurs, for example, on some Lappish (Saami) shamanic drums with a central solar symbols and radial-centric system of the image (Fig. 9).

The God Radien, his wife and son of Radien with his wife or sister are portrayed upside supreme at the top of these drums [52, p. 32, 35]. In the Saami mythology Radien - the supreme god. More archaic image of the supreme deity among the Scandinavian Saami - Veralden - radien (Veralden-Olmany) - "Man (lord) of the Universe". It was believed that he supports the vault of heaven and a tree or post - "support of the sky" placed over him altar [42, p. 462], which is in fact symbolized the World Tree.

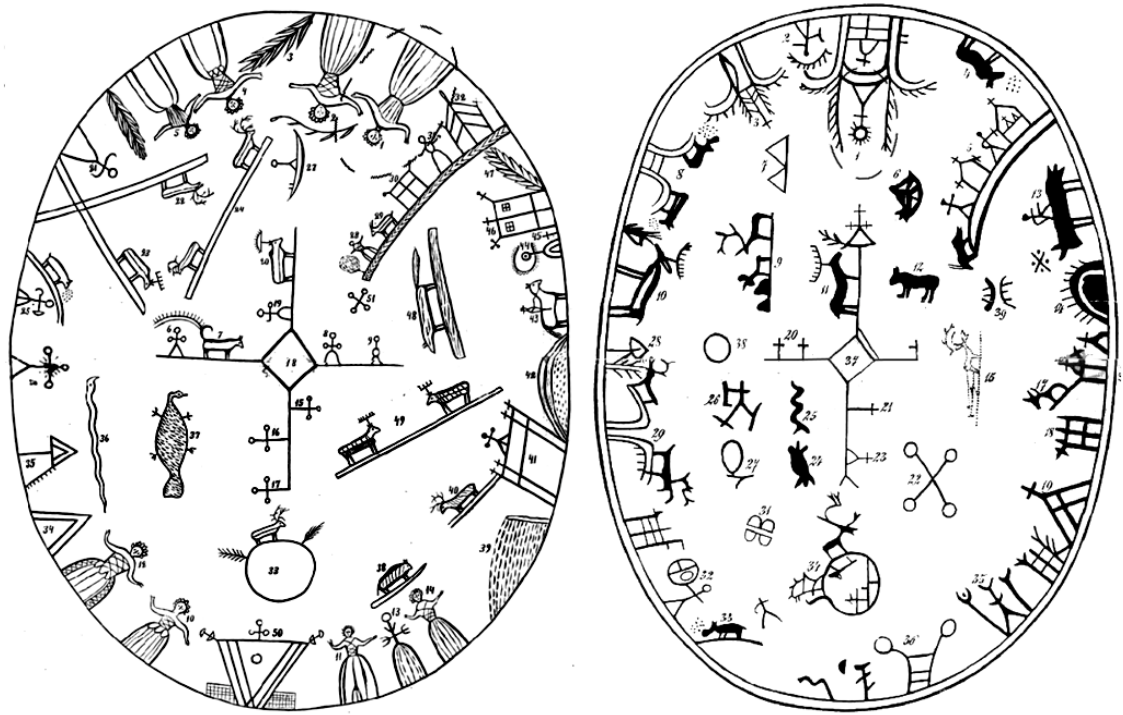

**Figure 9.** Lappish drums with radial-centric system images [50, Fig. 2, Figure 4]. Inverted images of Radiena are marked number one in the top of the drums. The authors singled by dotted line images of Radiena for clarity.

Thus, as a result of our study on Varvarinsky slab, identified by us as prototype of analemmatic sundial, petroglyph "tree" has been revealed, which performs additional functions markup. Mark up of ellipses of analemmatic sundial "dials" for northern latitudes was possible with it.

Comparative analysis of the images at Varvarinsky slab and Srubna vessels allowed us to determine the sign of the inverted tree, as a symbol denoting direction to the North and / or the North Pole of the World among Srubna population and may be among their predecessors; interpret some accompanying signs, as symbols of circumpolar stars and constellations (asterisms), and the god similarly Varuna - the main heavenly deity of the Vedic pantheon, identify as the supreme deity of Srubna pantheon.

The authors express their sincere gratitude for the support studies to T.F. Knyazeva, S.A. Husser, A.N. Usachuk, V.A. Larenok and employees of the department of archaeological heritage GOUK RO "Donskoe nasledie".

## List of Abbreviations

GOUK RO – Gosudarstvennoe Avtonomnoe Uchrezhdenie Kul'tury Rostovskoy Oblasti
[State autonomous organization of Culture of Rostov Region]

SOIGSI – Severo-Osetinskiy Institut Gumanitarnykh i Sotsial'nykh Issledovaniy [North
Ossetian Institute of Humanitarian and Social Research]

TAE – Taganrogskaya Arkheologicheskaya Ekspeditsiya [Taganrog Archaeological
Expedition]

TGLIAMZ – Taganrogskiy Gosudarstvennyy Literaturnyy i Istoriko- Arkhitekturnyy
Muzey-Zapovednik [Taganrog State Literary and Historico-Architectural Museum-Reserve]